\documentclass[a4paper,11pt]{article}
\pdfoutput=1 

\usepackage{jcappub} 
\usepackage{indentfirst}
\usepackage{appendix}
\usepackage{multirow}

\title{A preliminary forecast for cosmological parameter estimation with gravitational-wave standard sirens from TianQin}

\author[a]{Ling-Feng Wang,}
\author[a]{Ze-Wei Zhao,}
\author[a]{Jing-Fei Zhang,}
\author[a,b,c,1]{Xin Zhang\note{Corresponding author.}}

\affiliation[a]{Department of Physics, College of Sciences, Northeastern
University, Shenyang 110819, China}
\affiliation[b]{Ministry of Education's Key Laboratory of Data Analytics and Optimization
for Smart Industry, Northeastern University, Shenyang 110819, China}
\affiliation[c]{Center for High Energy Physics, Peking University, Beijing 100080, China}

\emailAdd{1810023@stu.neu.edu.cn, 1810024@stu.neu.edu.cn, jfzhang@mail.neu.edu.cn, zhangxin@mail.neu.edu.cn}

\abstract{TianQin is a space-based gravitational-wave observatory scheduled to be launched in the 2030s. In this work, we make a preliminary forecast for the cosmological parameter estimation with the gravitational-wave standard siren observation from TianQin. We simulate the standard siren data of TianQin based on its 5-year observation after the completion of construction. In the simulation, three models for the population of massive black hole binary (MBHB), i.e., pop III, Q3nod, and Q3d, are considered to predict the event numbers of MBHB mergers. We find that: (i) among the three MBHB models, the Q3nod model can provide the tightest constraints on the cosmological parameters; (ii) TianQin's standard siren observation can effectively break the parameter degeneracies inherent in the cosmic microwave background observation; and (iii) the future standard siren observation from TianQin can significantly improve the cosmological parameter estimation under the current mainstream electromagnetic observations.}

\begin{document}
\maketitle
\flushbottom

\section{Introduction}\label{sec:intro}

The observation of the binary neutron star (BNS) merger gravitational wave (GW) event GW170817 \cite{TheLIGOScientific:2017qsa} initiated the multi-messenger astronomy era, opening a new window to measure the expansion history of the universe. We can simultaneously determine the absolute luminosity distance to the event source by GW observation and the redshift of the source by optical observation. This will provide a new cosmological probe, known as ``standard sirens" \cite{Schutz1986,Holz:2005df}. Therefore, by the observation of standard sirens, the relationship between cosmic distances and redshifts could be established, which can be used to constrain cosmological models. A significant advantage of the standard siren observation is that it provides a measurement for the absolute luminosity distance that is calibrated only by theory, and independent of the complex astrophysical distance ladder with poorly understood calibration processes. From the actual standard siren observation of the BNS merger event (GW170817 and GRB 170817A) \cite{GBM:2017lvd,Monitor:2017mdv}, an independent measurement of the Hubble constant has been made \cite{Abbott:2017xzu}. It has been predicted that in the forthcoming years, the observation of BNS merger events as standard sirens from the ground-based detector network is expected to serve as an arbitration for the well-known tension of the Hubble constant in cosmology \cite{Feeney:2018mkj,Chen:2017rfc} (see also Ref.~\cite{Zhang:2019ylr}). The forecasts for using the GW standard sirens observed from the third-generation ground-based detectors to constrain cosmology have been recently intensively discussed; see, e.g., Refs.~\cite{Zhang:2019ylr,Zhang:2018byx,Wang:2018lun,Cai:2016sby,Cai:2017aea,Cai:2017buj,Sathyaprakash:2009xt,Zhao:2010sz,Li:2013lza,Yang:2017bkv,Wei:2018cov,Du:2018tia,Wei:2019fwp,Fu:2019oll,Yang:2019bpr,Cai:2019cfw,Nunes:2019bjq,Yang:2019vni,Mendonca:2019yfo,Zhang:2019ple,Zhang:2019loq,Li:2019ajo,Jin:2020hmc}. The standard sirens at much higher redshifts provided by the massive black hole binary (MBHB) coalescences will be observed by space-based GW interferometers, and these high-redshift observations will also play an important role in the cosmological parameter estimation.

\begin{figure}[h]
\begin{center}
\includegraphics[scale=0.13]{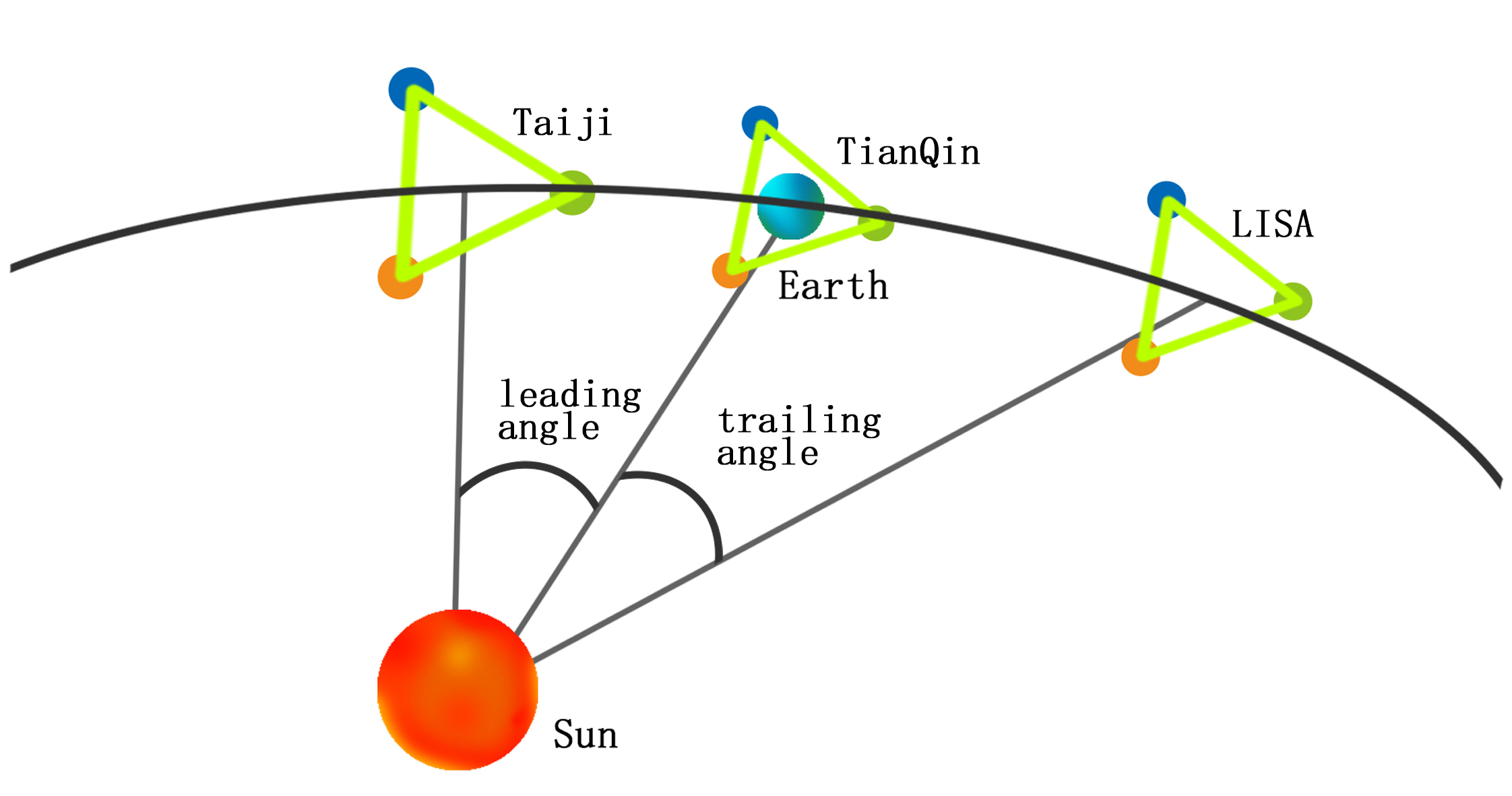}
\end{center}
\caption{The schematic diagram of the orbits of Taiji, LISA, and TianQin.} \label{orbit}
\end{figure}

The Laser Interferometer Space Antenna (LISA) \cite{LISA,Audley:2017drz,Armano:2016bkm,Armano:2018kix,Abich:2019cci} is an European space-based GW detector, aiming at detecting low-frequency GWs in the millihertz frequency range (0.1 mHz -- 1 Hz).
Recently, LISA Pathfinder \cite{Armano:2016bkm,Armano:2018kix} successfully demonstrated the drag-free technology, and the Gravity Recovery and Climate Experiment (GRACE) \cite{Abich:2019cci} further tested the laser metrology. LISA consists of three identical drag-free spacecrafts forming an equilateral triangular constellation with $2.5\times10^{6}$ km arm length. The plane of spacecrafts is at an angle of $60^\circ$ relative to the ecliptic plane. LISA is considered to be located behind the Earth and follows it in the ecliptic with $20^{\circ}$ trailing angle \cite{Audley:2017drz}.
There are also two space projects proposed by Chinese researchers, namely, TianQin \cite{Mei:2015joa,Luo:2015ght,Hu:2018yqb,Feng:2019wgq,Wang:2019ryf,Shi:2019hqa,Mei:2020lrl} and Taiji \cite{Wu:2018clg,Wang:2017aqq,Guo:2018npi,Wang:2020vkg,Hu:2017mde,Luo:2019RIP,Ruan:2020smc}. The Taiji mission is a LISA-like space-based GW detector in heliocentric orbit with $3\times10^{6}$ km arm length. Taiji-1, the pathfinder of Taiji, was launched in August 2019 \cite{Luo:2019RIP} to test the first stage of development of space technologies. In order to make Taiji and LISA be separated relatively far, Taiji is planned to be localized in front of the Earth with $20^{\circ}$ leading angle, such that the LISA-Taiji network \cite{Ruan:2020smc,Wang:2020vkg} could have a powerful joint detection capability. The schematic diagram of the orbits of Taiji, LISA, and TianQin is shown in Fig.~\ref{orbit}.

TianQin project is a space-based GW detector with $1.7\times 10^5$ km arm length in the geocentric orbit, and is planned to run in a detector mode to detect a known reference source firstly. In December 2019, TianQin-1, the first pathfinder satellite, was successfully launched and tested the first stage of development of their technologies in space. This implies that the key technologies of TianQin are gradually maturing, paving the way for its launch in the 2030s. The normal direction of the plane of the TianQin's spacecrafts points towards the specific reference source RX J0806.3+1527 during its operation. Because the J0806.3+1527 locates in $\sim 4.7^\circ$ from the ecliptic plane, the detector plane stands nearly vertical to the ecliptic plane \cite{Luo:2015ght}. In order to maintain the thermal stability of the detector, TianQin has two observation windows in one-year operation, resulting in the ``3 month on + 3 month off" working pattern of TianQin, which would somehow inevitably limit its performance in cosmology. Thus we adopt the twin-constellation scenario with two sets of perpendicular constellation operating in succession, which would provide a complete time coverage and double the detection rate of GW events \cite{Wang:2019ryf}. LISA, Taiji, and TianQin have various configurations and thus result in different impacts on the future cosmological parameter estimation. 
Recently, some studies on the capability of LISA and Taiji in the aspect of improving the cosmological parameter estimation have been seriously conducted \cite{Tamanini:2016zlh,Belgacem:2019pkk,Zhao:2019gyk}. Nevertheless, the forecast study on the prospect of TianQin in the future cosmological parameter estimation is still absent to date. Thus, it is of great importance to assess the capability for the TianQin project in the cosmological parameter estimation by using the simulated standard siren data.

Recently, the Event Horizon Telescope captured the image of a massive black hole (MBH) in the center of M87 \cite{Akiyama:2019cqa}, directly showing the existence of MBH in the center of galaxy. (See Refs.~\cite{Tsupko:2019pzg,Qi:2019zdk} for discussions on a new cosmological probe from supermassive black holes.) The inspiral, merger, and ringdown of MBHBs may all be detected by the space-based GW interferometers. The certain formation mechanism of MBHB is still unclear, but it is pointed out that two factors can mainly influence the predicted number of the observable MBHB merger events, i.e., the MBHs seeding at high redshifts, and the delay between the merger of two MBHs and that of their host galaxies. We consider three models for the population of MBHB in this work, and we will discuss them in more detail in the following section.

The rest of this paper is organized as follows. In Sec. \ref{sec:Method}, we describe the cosmological models considered in this work and introduce the simulation method for the observational data of TianQin, and we also explain the method for constraining cosmological parameters. The results are shown and discussed in Sec. \ref{sec:Result}. The conclusion is given in Sec. \ref{sec:con}. Unless otherwise specified we shall adopt the system of units in which $c=G=1$ throughout this paper.

\section{Methods and data}\label{sec:Method}

\subsection{Methods of simulating the TianQin's standard siren data}

\subsubsection{The configuration of TianQin}\label{subsubsection:A1}
We briefly introduce the relevant information of TianQin and describe the methods of simulating its standard siren data.
The GW strain $h(t)$ can be described by two independent polarizations ${h_{+,\times}}(t)$ in the transverse-traceless gauge,
\begin{align}
 h(t) = {F_ + }(t;\theta ,\phi ,\psi ){h_ + }(t) + {F_ \times }(t;\theta ,\phi ,\psi ){h_ \times }(t)  \,,
\end{align}
where $F_{+,\times}$ are antenna pattern functions, $(\theta,\phi )$ denote source's polar angle and azimuthal angle in the ecliptic frame, and $\psi $ is the polarization angle of GW. We can separate the antenna pattern function into a polarization angle part and a $D_{+,\times}$ part that describes the dependence of time,
\begin{align}
 F_{+}(t) = {D_ + }(t,f)\cos (2\psi ) - {D_ \times }(t,f)\sin (2\psi )  \,, \label{Fplus}\\
 F_{\times }(t) = {D_ + }(t,f)\sin (2\psi ) + {D_ \times }(t,f)\cos (2\psi )  \,.\label{Fcros}
\end{align}
The forms of $D_{+,\times}$ rely on the specific configurations (e.g., orbit and orientation) of a GW detector and generally depend on the frequency of GWs.
For the inspiral phase of MBHBs, by adopting the low-frequency limit, $D_{+,\times}$ could be independent of frequency and written as \cite{Feng:2019wgq}
\begin{align}
  {D_+}(t)=&\frac{{\sqrt 3 }}{{32}}\bigg(4\cos (2{\kappa _1(t)})\Big(\big(3 + \cos(2\theta)\big)\cos{\bar \theta}\sin(2\phi - 2{\bar \phi}) + 2\sin(\phi - {\bar \phi})\sin(2\theta)\sin {\bar \theta} \Big) \nonumber\\
  &-\sin(2{\kappa _1(t)}) \Big(3 + \cos(2\phi - 2{\bar \phi})\big(9 + \cos(2\theta)(3 + \cos(2{\bar \theta}))\big) - 6\cos(2{\bar \theta}) \nonumber\\
  &\times\sin^2(\phi - {\bar \phi}) - 6\cos(2\theta)\sin^2 {\bar \theta} + 4\cos(\phi - {\bar \phi})\sin(2\theta)\sin(2{\bar \theta})\Big)\bigg)  \,,\label{Dplus}  \\
  {D_\times}(t) =& \frac{{\sqrt 3 }}{8}\bigg(-4\cos(2{\kappa _1(t)})\Big(\cos(2\phi - 2{\bar \phi})\cos \theta \cos {\bar \theta} + \cos(\phi - {\bar \phi})\sin \theta \sin {\bar \theta} \Big) + \sin(2{\kappa _1(t)}) \nonumber\\
  &\times\Big(-\cos \theta \big(3 + \cos(2{\bar \theta})\big)\sin(2\phi - 2{\bar \phi})
   - 2\sin(\phi - {\bar \phi}) \sin \theta \sin(2{\bar \theta})\Big)\bigg)  \,,\label{Dcros}
\end{align}
where $\kappa _{1}(t) = 2\pi f_{\rm{sc}}t + \kappa _0$, with $f_{\rm{sc}}=1/(3.65$ day)$=3.17\times 10^{-6}$ Hz representing the rotation frequency of spacecrafts around the Earth, and $\kappa _0$ is a constant phase term depending on the setup of satellites' coordinates. Here we choose $\kappa _0=0$ for simplicity. $(\bar \theta = 1.65, \bar \phi = 2.10)$ describe the polar angle and the azimuthal angle of the reference source RX J0806.3+1527 in the heliocentric-ecliptic frame.
Using Eqs.~(\ref{Fplus})--(\ref{Dcros}), we can derive the antenna pattern function of an equivalent detector of TianQin,
$F_{ + ,\times}^{(1)}(t;\theta ,\phi ,\psi)$, and another equivalent detector's antenna pattern function is just \cite{Shi:2019hqa}
\begin{eqnarray}
F_{+ ,\times}^{(2)}(t;\theta ,\phi ,\psi)=F_{+ ,\times}^{(1)}(t;\theta ,\phi-\pi/4 ,\psi ).
\end{eqnarray}

By using the stationary phase approximation under required constraints \cite{Feng:2019wgq} and replacing $t$ in Eqs.~(\ref{Dplus})--(\ref{Dcros}) by $t(f) = {t_{\rm c}} - \frac{5}{256} M_{{\rm c}}^{-5/3}(\pi f)^{-8/3}$ \cite{Krolak:1995md,Buonanno:2009zt}, where $t_{\rm c}$ is the coalescence time and is set to be zero in our analysis, the Fourier transformation of the strain can be obtained,
\begin{eqnarray}
  \tilde h(f) = AQf^{-7/6}{e^{i \Psi (f)}}, \quad  {\rm{for}} ~ f>0  \,,\label{eq:hf}
\end{eqnarray}
where
\begin{equation} \label{eq:amp}
 A =  - \sqrt {\frac{5}{96}} \frac {{M_{\rm c}}^{5 /6}} {\pi ^{2/3}{d_{\rm L}}}  \,,
\end{equation}
and
\begin{equation}\label{eq:Qfac}
  Q = \sqrt{ (1 + \cos^2{\iota})^2 F^2_{+}\big(t(f)\big) + (2\cos\iota)^2 F^2_{\times} \big(t(f)\big) }  \,.
\end{equation}
Here, we define $M_{\rm c} = (1 + z)\eta^{3/5}M$ as the redshifted chirp mass observed in the reference frame of detectors with the total mass $M = {m_1} + {m_2}$ and the symmetric mass ratio ${\eta  = m_1m_2 /M^2 }$. The GW strain phase evolution $\Psi(f)$ here will be eliminated in our calculations, as shown in Eq.~(\ref{induct}) below.

Besides, $d_{\rm L}$ is the luminosity distance at redshift $z$ in a given cosmological model,
\begin{align}\label{eq:dL0}
d_{\rm L}(z)=\frac{1+z}{H_{0}}\int_{0}^{z}\frac{dz^{\prime}}{E(z^{\prime})},
\end{align}
where $E(z)\equiv H(z)/H_{0}$, and $H_{0}=100h~{\rm km}~{\rm s}^{-1}~{\rm Mpc}^{-1}$ is the Hubble constant. The Hubble parameter $H(z)$ can be written as
\begin{align}
H^2(z) = &~H_0^2 \left\{ (1 - {\Omega _m})\exp \left[3\int_0^z {\frac{{1 + w(z')}}{{1 + z'}}} d z'\right]\right.+{\Omega _m}{(1 + z)^3}\bigg\}.
\label{equa:H}
\end{align}
Here $\Omega _m$ is the current matter density parameter, and the parameter $w(z)=p_{\rm de}(z)/\rho_{\rm de}(z)$ describes the equation of state (EoS) of dark energy. The $\Lambda$CDM model [$w(z) = -1$] and the $w$CDM model [$w(z) = \rm{constant}$] are considered in this paper.

\begin{figure}[h]
\begin{center}
\includegraphics[scale=0.65]{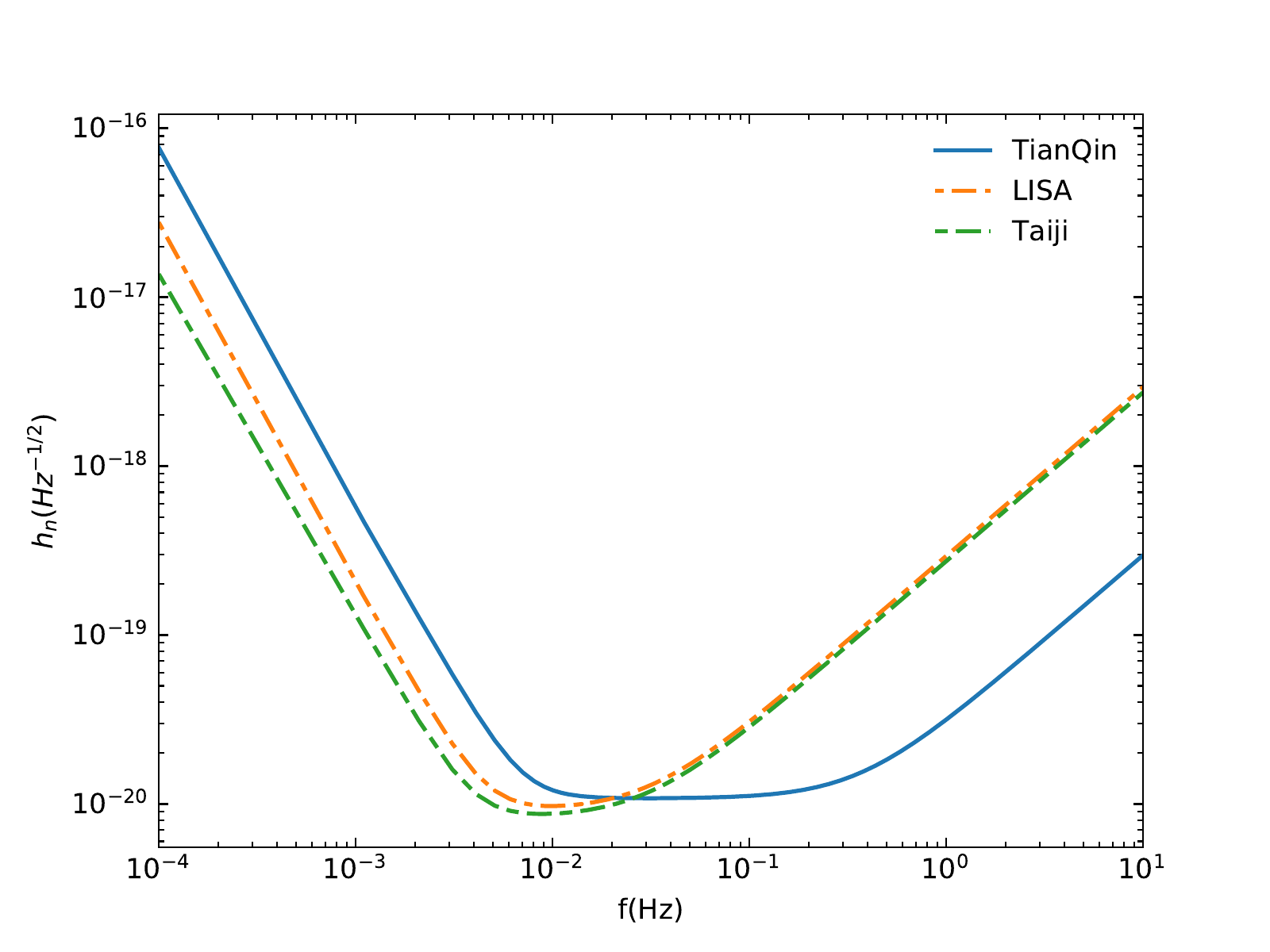}
\end{center}
\caption{The sensitivity curves of TianQin, Taiji, and LISA with the
configuration N2A2M5L6.} \label{Sn}
\end{figure}

With the recent configuration of TianQin, the one-sided noise power spectral density (PSD) is given by \cite{Feng:2019wgq,Wang:2019ryf,Shi:2019hqa}
\begin{align}
\label{curveTianQin}
S_n(f)=\Big[\frac{4S_a}{(2\pi f)^4L_0^2}\Big(1+\frac{10^{-4}{\rm Hz}}{f}\Big) +\frac{S_x(f)}{L_0^2}\Big]\times\Big[1+\Big(\frac{2fL_0/c}{0.41}\Big)^2\Big],
\end{align}
where
$L_{0} = \sqrt{3}\times 10^8~{\rm m}$ is the arm length, $c$ is the speed of light, $S_{x} = 10^{- 24}~{\rm{m}}^{2}{\rm{Hz}}^{-1}$, and
$S_{a} = 10^{- 30}~{\rm{m}}^{2}{\rm{s}}^{- 4}{\rm{Hz}}^{-1}$
are the PSDs of position noise and residual acceleration noise, respectively. The amplitude spectral density (ASD) of effective strain noise is defined by $h_{n}(f)=\sqrt{S_{n}(f)}$. The sensitivity curve of TianQin is shown in Fig.~\ref{Sn}. In order to make a comparison, we also show the sensitivity curves of LISA (N2A2M5L6)\footnote{The ``N2A2M5L6" represents the configuration of LISA with the noise level ``N2", the arm length of $2\times 10^9$m (A2), a five-year mission lifetime (M5), and six links (L6).} \cite{Tamanini:2016zlh} and Taiji in the same figure.

The combined signal-to-noise ratio (SNR) for the network of two equivalent independent interferometers is
\begin{equation}
\rho=\sqrt{\sum\limits_{i=1}^{2}(\rho^{(i)})^2},
\label{euqa:rho}
\end{equation}
where $\rho^{(i)}=\sqrt{({h}^{(i)}|{h}^{(i)})}$, with the inner product being defined as
\begin{align}
\label{induct}
(a|b)\equiv4\int_{f_{\rm low}}^{f_{\rm up}}\frac{\tilde{a}(f)\tilde{b}^{*}(f)+\tilde{a}^{*}(f)\tilde{b}(f)}{2}\frac{df}{S_{n}(f)},
\end{align}
where ``$\sim$" above a function denotes the Fourier transform of the function.
We choose $f_{\rm low}=10^{-4}\,\rm Hz$ as a conservative lower frequency cutoff,
and $f_{\rm up}$ is set to be the innermost stable circular orbit (ISCO) frequency $f_{\rm ISCO}=c^3/(6\sqrt{6}\pi G M)$ ~\cite{Feng:2019wgq}. Since the analytic fit of the noise power spectrum $S_{n}(f)$ assumes the long-wavelength approximation, which may break down at $f_{\rm max}=c/(2\pi L_{0})\simeq0.05\frac{\rm Gm}{L_{0}}\,\rm Hz$ \cite{Klein:2015hvg}, we take this frequency as the maximum value for the upper limit of the integral.

\subsubsection{The properties of GW sources and detection rates}\label{subsubsection:A2}
We then discuss the source properties of standard sirens. As mentioned in Sec.~\ref{sec:intro}, two factors determine the classification of the MBHB models: the seeds of MBHs, and the delay between the merger of MBHs and that of their host galaxies. Due to the uncertainty of the birth mechanism of MBHs, MBHs are divided into two scenarios, namely the ``light-seed" scenario and the ``heavy-seed" scenario.
Based on these two factors above, there are three proposed models for the population of MBHB \cite{Klein:2015hvg}.

(1) Model pop III: a ``realistic" light-seed model including delay, representing that the MBHs grow from the remnants of population III (pop III) stars.

(2) Model Q3d: a ``realistic" heavy-seed model including delay, representing that the MBHs grow from the collapse of protogalactic disks.

(3) Model Q3nod: the same as model Q3d, but without delay, as an ``optimistic" scenario for the predicted event rates.

For the mass of MBHs, a flat distribution function within the interval [$10^{4}$, $10^{7}$] $M_\odot$ is considered. Compared with the mass distribution functions considered in Ref.~\cite{Klein:2015hvg}, this is an approximation for the calculation of SNR and will affect the accuracy of instrumental error estimation. In the practical calculation, this approximation is acceptable for a preliminary estimation as there are other important errors mentioned below that will affect the final results, such as the weak-lensing error. The redshift distributions of MBHs are chosen to be consistent with \textcolor{blue}{Ref.~\cite{Tamanini:2016uin}}. We randomly sample the position angle $\theta$, $\phi$, and the polarization angle $\psi$ of MBHs in the parameter intervals [0, $\pi$], [0, 2$\pi$], and [0, $\pi$], respectively.

The detection rates of TianQin for the three MBHB models are given in Ref.~\cite{Wang:2019ryf}. For the twin-constellation senario, the detection rates for the models of pop III, Q3d, and Q3nod are approximately 23 ${\rm yr}^{-1}$, 8 ${\rm yr}^{-1}$, and 118 ${\rm yr}^{-1}$, respectively. Hence, within a 5-year observation, the numbers of detected MBHB merger events by TianQin are about 115 for pop III model, 40 for Q3d model, and 590 for Q3nod model. As shown in Ref.~\cite{Tamanini:2016zlh}, for the configuration N1A2M5L6 of LISA, it has a similar detection capability compared with TianQin. There would be approximately $10 \%$ (pop III), $35 \%$ (Q3d), and $4.5 \%$ (Q3nod) GW events whose electromagnetic (EM) counterparts could be detected by the future projects such as the Square Kilometre Array (SKA) \cite{SKA}, the Large Synoptic Survey Telescope (LSST) \cite{LSST}, and the Extremely Large Telescope (ELT) \cite{ELT}. By applying these percentages to the event numbers of TianQin, 12, 14, and 27 standard siren events are considered in our simulation for the models of pop III, Q3d, and Q3nod, respectively.

\subsection{Methods of constraining cosmological parameters}
We construct 100 catalogs for each MBHB model and each cosmological model. The Fisher matrix technique is used to evaluate the errors of cosmological parameters for a given catalog. For a cosmological model with parameter $\theta_i$, the entry of the Fisher matrix is defined as
\begin{align}
F_{ij}=\sum_n \frac{1}{(\sigma_{d_{\rm L}})^2(z_n)} \frac{\partial d_{\rm L}(z_n)}{\partial \theta_i}\bigg|_{\rm fid} \frac{\partial d_{\rm L}(z_n)}{\partial \theta_j}\bigg|_{\rm fid},
\end{align}
where the sum is over all MBHB merger events in a given catalog, $z_n$ is the redshift of the $n$th GW event, and the derivatives of $d_{\rm L}$ are evaluated at the fiducial values of cosmological models. The fiducial values of the cosmological parameters used in this paper are the constraint results from the Planck 2015 observation \cite{Ade:2015xua}. The total measurement error of luminosity distance $\sigma_{d_{\rm L}}$ consists of the lensing error, the instrumental error, the peculiar velocity error, and the redshift measurement error, which can be expressed as \begin{align}\label{eq:DECIGOlensingError}
(\sigma_{d_{\rm L}})^{2}=(\sigma_{d_{\rm L}}^{\rm lens})^{2}+(\sigma_{d_{\rm L}}^{\rm inst})^{2}+(\sigma_{d_{\rm L}}^{\rm pv})^{2}+(\sigma_{d_{\rm L}}^{\rm reds})^{2}.
\end{align}

For the measurement error of luminosity distance $\sigma_{d_{\rm L}}$, the main systematic error comes from weak-lensing, especially at high redshifts. We adopt the weak-lensing error from the fitting formula \cite{Tamanini:2016zlh},
\begin{align}\label{eq:DECIGOlensingError}
\sigma_{d_{\rm L}}^{\rm lens}(z)=d_{\rm L}(z)\times 0.066\bigg[\frac{1-(1+z)^{-0.25}}{0.25}\bigg]^{1.8}.
\end{align}

By applying the Fisher matrix to the waveform, it can be found that $\sigma_{d_{\rm L}}^{\rm inst}\simeq d_{\rm L}/{\rho}$. Considering the correlation between $d_{\rm L}$ and inclination $\iota$ \cite{Cai:2017aea}, a factor of 2 should be added into the instrumental error, i.e.,
\begin{align}
\sigma_{d_{\rm L}}^{\rm inst}\simeq \frac{2d_{\rm L}}{\rho}.
\end{align}

The error caused by the peculiar velocities of sources should also be included,
\begin{align}
\sigma_{d_{\rm L}}^{\rm pv}(z)=d_{\rm L}(z)\times\bigg[1+\frac{c(1+z)^2}{H(z)d_{\rm L}(z)}\bigg]\frac{\sqrt{\langle v^{2}\rangle}}{c},
\end{align}
where the peculiar velocity $\sqrt{\langle v^{2}\rangle}$ of the source with respect to the Hubble flow is roughly set to be $500\,\mathrm{km\,s^{-1}}$.

The error from the redshift measurement of the EM counterpart could be ignored if the redshift is measured spectroscopically. But when using photometric redshift for the distant source, this factor should be taken into account. For this reason, we estimate the error on the redshift measurement as
\begin{align}
\sigma_{d_{\rm L}}^{\rm reds}=\frac{\partial d_{\rm L}}{\partial z} (\Delta z)_n,
\end{align}
with $(\Delta z)_n\simeq 0.03(1+z_n)$ \cite{Ilbert:2013bf}. In principle, the spectroscopic redshifts are expected to be uncertain at $z>2$ \cite{Dahlen:2013fea}. Nevertheless, as an optimistic forecast, we also assume these measurements for high-redshift events. We consider 4, 5, and 12 photometric observation events for the pop III, Q3d, and Q3nod models, respectively, according to the proportion of the photometric observation events provided in Ref.~\cite{Tamanini:2016zlh}.

\begin{figure}[h]
\begin{center}
\includegraphics[scale=0.65]{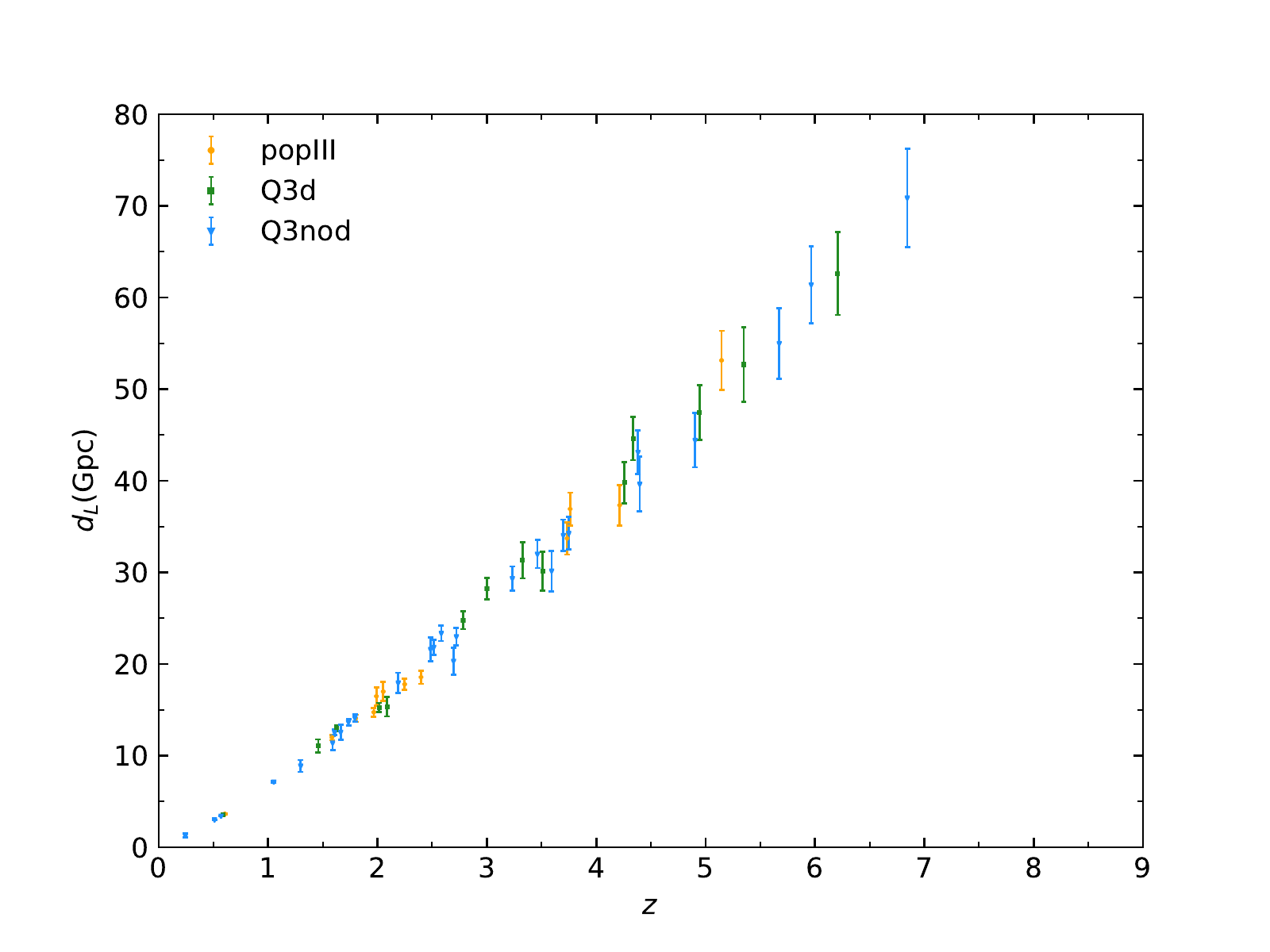}
\end{center}
\caption{The simulated standard sirens of TianQin during its 5-year operation based on the $\Lambda$CDM cosmology. The redshift distributions and the measurement errors of luminosity distances for the three models for the population of MBHB are shown.} \label{sirens}
\end{figure}

We calculate the average value of the Fisher matrix of the 100 catalogs, and select the catalog that gives the value close to the mean as a representative to perform the Markov-chain Monte Carlo (MCMC) analysis \cite{Lewis:2002ah}. The standard siren data simulated from the selected catalogs are shown in Fig.~\ref{sirens}.
For the GW standard siren measurement with $N$ simulated data points, we can write its $\chi^2$ as
\begin{align}
\chi_{\rm GW}^2=\sum\limits_{i=1}^{N}\left[\frac{\bar{d}_{\rm L}^i-d_{\rm L}(\bar{z}_i;\vec{\Omega})}{\bar{\sigma}_{d_{\rm L}}^i}\right]^2,
\label{equa:chi2}
\end{align}
where $\bar{z}_i$, $\bar{d}_{\rm L}^i$, and $\bar{\sigma}_{d_{\rm L}}^i$ are the $i$th redshift, luminosity distance, and error of luminosity distance for the simulated GW data, and $\vec{\Omega}$ represents the set of cosmological parameters.

We will make a comparison for the current EM observational data with the GW standard siren data from TianQin, so as to further discuss how the TianQin can improve the constraints on various cosmological parameters. For the current cosmological observations, we consider three mainstream data sets at present, i.e., cosmic microwave background (CMB) anisotropies, baryon acoustic oscillations (BAO), and type Ia supernovae (SN).
For the CMB data, we employ the ``Planck distance priors'' from the Planck 2018 observation \cite{Ade:2015rim}, which is considerably sufficient to explore the cosmic expansion history in this work. For the BAO data, we use four data points from the six-degree-field galaxy survey (6dFGS) at $z_{\rm eff}=0.106$ \cite{Beutler:2011hx}, the SDSS main galaxy sample (MGS) at $z_{\rm eff}=0.15$ \cite{Ross:2014qpa}, the baryon oscillation spectroscopic survey (BOSS) LOWZ at $z_{\rm eff}=0.32$ \cite{Anderson:2013zyy}, and the BOSS CMASS at $z_{\rm eff}=0.57$ \cite{Anderson:2013zyy}. For the SN data, we use the latest sample from the the Pantheon compilation \cite{Scolnic:2017caz}.

\section{Results and discussion}\label{sec:Result}

\begin{table}
\caption{Constraint errors (1$\sigma$ level) and accuracies for the cosmological parameters in the $\Lambda$CDM model, by using TQ, CMB, CMB+TQ, CBS, and CBS+TQ data combinations. Here, TQ stands for the TianQin mock data, and CBS stands for the CMB+BAO+SN data combination.  The three values corresponding to TQ from top to bottom represent the models of pop III, Q3nod, and Q3d, respectively.}
\label{tab:LCDM}
\hspace{-0.5cm}
\begin{center}
{\centerline{\begin{tabular}{ccccccccc}
\hline
Model & \multicolumn{5}{c}{$\Lambda$CDM}\\
\cline{2-6}
Data & TQ & CMB & CMB+TQ & CBS & CBS+TQ\\
\hline
  & 0.0325 & \text{} & 0.0071 & \text{} & 0.0057\\
  $\sigma(\Omega_{\rm m})$& 0.0210 & 0.0085 & 0.0058 & 0.0063 & 0.0050\\
  & 0.0310 & \text{} & 0.0071 & \text{} & 0.0056\\
\hline
  & 0.0130 & \text{} & 0.0050 & \text{} & 0.0040\\
  $\sigma(h)$& 0.0076 & 0.0061 & 0.0041 & 0.0045 & 0.0035\\
  & 0.0120 & \text{} & 0.0050 & \text{} & 0.0040\\
\hline
  & 0.1006 & \text{} & 0.0223 & \text{} & 0.0179\\
  $\varepsilon(\Omega_{\rm m})$& 0.0658 & 0.0269 & 0.0181 & 0.0198 & 0.0155\\
  & 0.0963 & \text{} & 0.0223 & \text{} & 0.0177\\
\hline
  & 0.0194 & \text{} & 0.0074 & \text{} & 0.0060\\
  $\varepsilon(h)$& 0.0113 & 0.0090 & 0.0061 & 0.0067 & 0.0051\\
  & 0.0179 & \text{} & 0.0074 & \text{} & 0.0060\\
\hline
\end{tabular}}}
\end{center}
\end{table}

\begin{table}
\caption{Constraint errors (1$\sigma$ level) and accuracies on the cosmological parameters in the $w$CDM model, by using TQ, CMB, CMB+TQ, CBS, and CBS+TQ data combinations. Here, TQ stands for the TianQin mock data, and CBS stands for the CMB+BAO+SN data combination. The three values corresponding to TQ from top to bottom represent the models of pop III, Q3nod, and Q3d, respectively.}
\hspace{-0.5cm}
\begin{center}
{\centerline{\begin{tabular}{ccccccccc}
\hline
Model & \multicolumn{5}{c}{$w$CDM}\\
\cline{2-6}
Data & TQ & CMB & CMB+TQ & CBS & CBS+TQ\\
\hline
  & 0.0335 & \text{} & 0.0083 & \text{} & 0.0060\\
  $\sigma(\Omega_{\rm m})$& 0.0230 & 0.0650 & 0.0064 & 0.0087 & 0.0052\\
  & 0.0310 & \text{} & 0.0081 & \text{} & 0.0060\\
\hline
  & 0.0400 & \text{} & 0.0091 & \text{} & 0.0066\\
  $\sigma(h)$& 0.0175 & 0.0690 & 0.0071 & 0.0093 & 0.0057\\
  & 0.0390 & \text{} & 0.0090 & \text{} & 0.0066\\
\hline
  & 0.4950 & \text{} & 0.0420 & \text{} & 0.0331\\
  $\sigma(w)$& 0.2250 & 0.2350 & 0.0365 & 0.0386 & 0.0311\\
  & 0.4750 & \text{} & 0.0415 & \text{} & 0.0329\\
\hline
  & 0.1095 & \text{} & 0.0262 & \text{} & 0.0191\\
  $\varepsilon(\Omega_{\rm m})$& 0.0737 & 0.2006 & 0.0202 & 0.0277 & 0.0164\\
  & 0.1013 & \text{} & 0.0256 & \text{} & 0.0190\\
\hline
  & 0.0564 & \text{} & 0.0135 & \text{} & 0.0098\\
  $\varepsilon(h)$& 0.0257 & 0.1021 & 0.0105 & 0.0137 & 0.0085\\
  & 0.0549 & \text{} & 0.0133 & \text{} & 0.0097\\
\hline
  & 0.3536 & \text{} & 0.0420 & \text{} & 0.0325\\
  $\varepsilon(w)$& 0.2064 & 0.2350 & 0.0364 & 0.0377 & 0.0304\\
  & 0.3345 & \text{} & 0.0414 & \text{} & 0.0322\\
\hline
\end{tabular}}}
\end{center}
\label{tab:wCDM}
\end{table}


\begin{figure*}[htb]
\begin{center}
\includegraphics[width=\linewidth,angle=0]{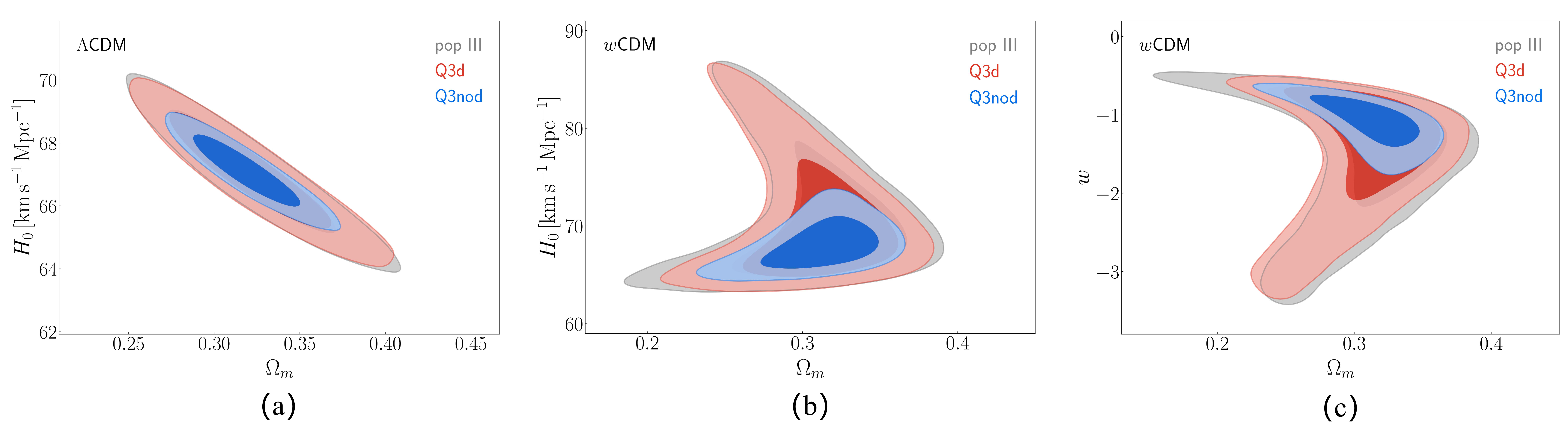}
\end{center}
\caption{Two-dimensional marginalized contours (68.3\% and 95.4\% confidence level) in the $\Omega_{m}$--$H_{0}$ plane for the $\Lambda$CDM model, in the $\Omega_{m}$--$H_{0}$ plane and $\Omega_{m}$--$w$ plane for the $w$CDM model by using the TianQin mock data alone. Three MBHB models are denoted by three different colors, i.e., grey (pop III), red (Q3d), and blue (Q3nod).} \label{TQ3models}
\end{figure*}

\begin{figure*}[htb]
\begin{center}
\includegraphics[width=\linewidth,angle=0]{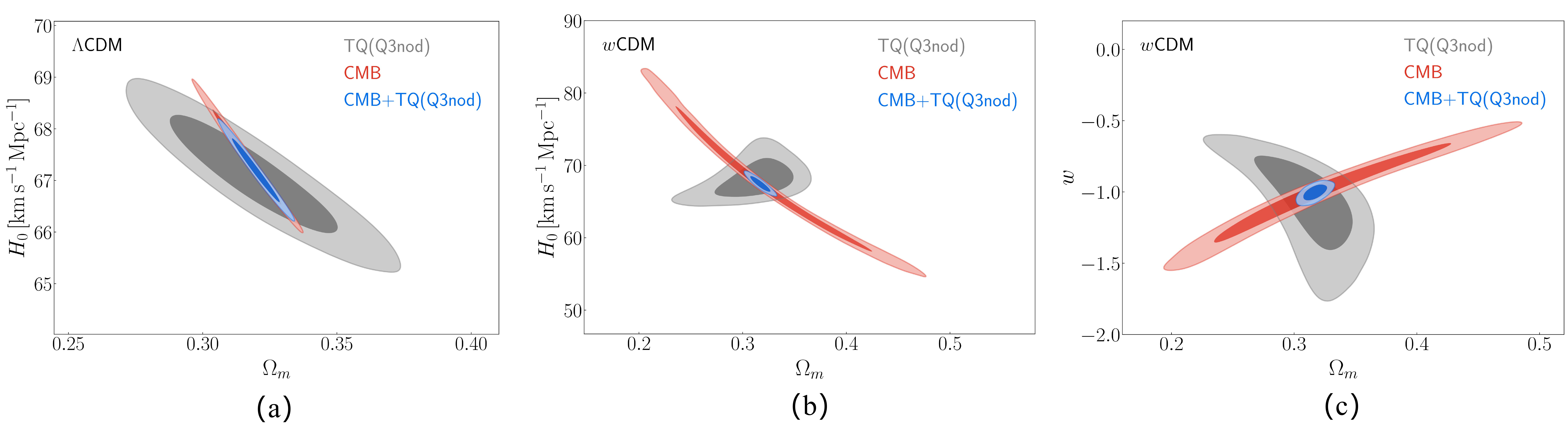}
\end{center}
\caption{Two-dimensional marginalized contours (68.3\% and 95.4\% confidence level) in the $\Omega_{m}$--$H_{0}$ plane for the $\Lambda$CDM model, in the $\Omega_{m}$--$H_{0}$ plane and $\Omega_{m}$--$w$ plane for the $w$CDM model by using the TianQin, CMB, and CMB+TianQin. Here, the TianQin mock data are simulated based on the Q3nod model.} \label{TQCMB}
\end{figure*}

\begin{figure*}[htb]
\begin{center}
\includegraphics[width=\linewidth,angle=0]{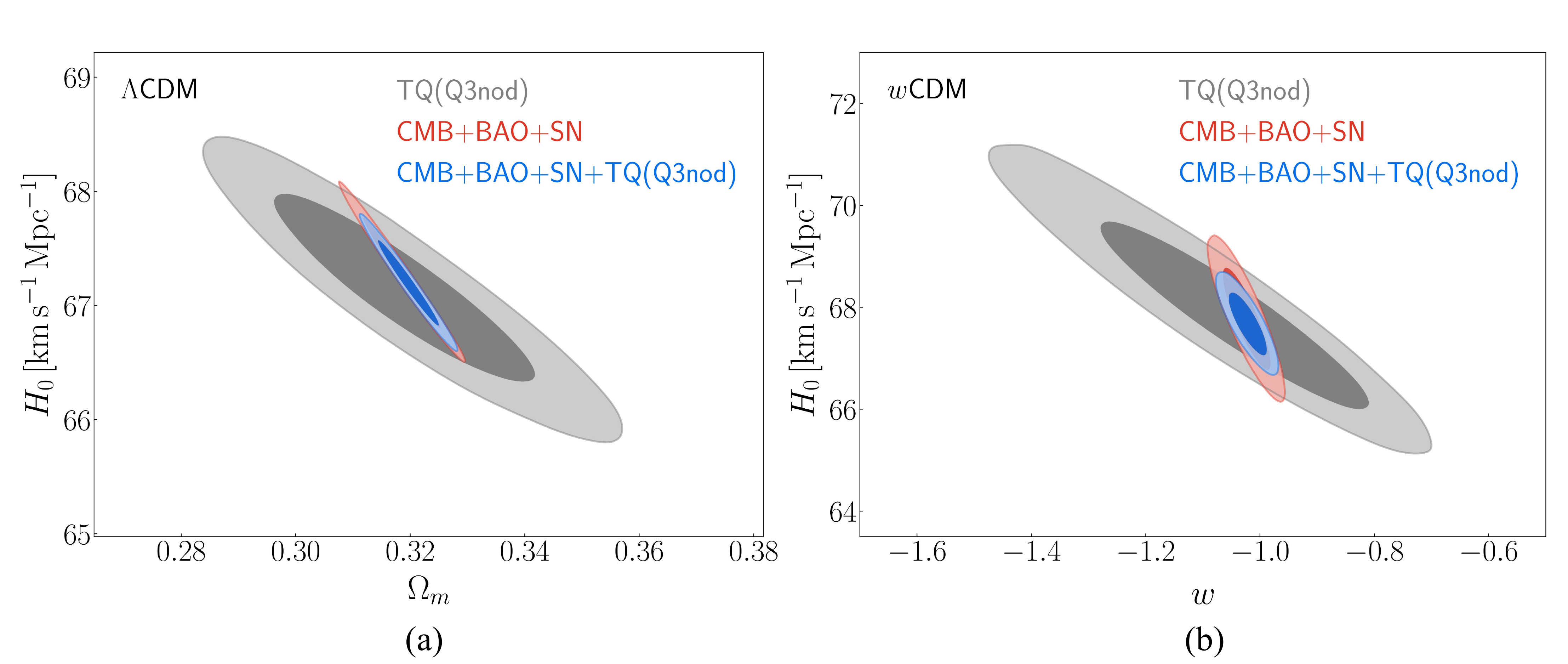}
\end{center}
\caption{Two-dimensional marginalized contours (68.3\% and 95.4\% confidence level) in the $\Omega_{m}$--$H_{0}$ plane for the $\Lambda$CDM model, in the $w$--$H_{0}$ plane for the $w$CDM model by using the TianQin, the CMB+BAO+SN, and the CMB+BAO+SN+TianQin data combinations. Here, the TianQin mock data are simulated based on the Q3nod model.} \label{TQCBS}
\end{figure*}

In this section, we shall report the constraint results for the considered cosmological models and make some relevant discussions on them.
First, we will show the constraints only using the standard siren data from TianQin. Next, we will show the constraint results by using the data combination of CMB+TianQin, and make an analysis on breaking the degeneracies inherent in CMB by TianQin. Finally, we will show the constraint results with the data combination of CMB+BAO+SN+TianQin, and make some discussions on the capability of TianQin in improving the constraint precision of the current mainstream observations.

The constraint results are given in Tables~\ref{tab:LCDM}--\ref{tab:wCDM}, and displayed in Figs.~\ref{TQ3models}--\ref{TQCBS}, in which we use the abbreviations ``TQ" and ``CBS" for convenience to represent the TianQin mock data and the CMB+BAO+SN data combination, respectively. In Tables~\ref{tab:LCDM}--\ref{tab:wCDM}, we list the $1\sigma$ constraint error and the constraint precision for each cosmological parameter. For a cosmological parameter $\xi$, we use $\sigma(\xi)$ and $\varepsilon(\xi)$ to denote its constraint error and precision, respectively.

Figure~\ref{TQ3models} shows the marginalized posterior probability distribution contours in the $\Omega_{\rm m}$--$H_{0}$ plane for the $\Lambda$CDM model (a), as well as in the $\Omega_{\rm m}$--$H_{0}$ plane (b) and in the $\Omega_{\rm m}$--$w$ plane (c) for the $w$CDM model, under the constraints of the simulated standard siren data from TianQin. In this figure, the TianQin mock data are simulated based on the three different MBHB models, which are marked with the colors of grey (pop III), red (Q3d), and blue (Q3nod).

We find that, among these three MBHB models, the Q3nod model could provide the tightest constraints on not only the $\Lambda$CDM but also the $w$CDM cosmological models. Besides, the pop III case and the Q3d case would provide similar constraints on both of these two cosmological models. This is because that the predicted number of standard sirens in the Q3nod model is 27, wheras the predicted numbers in the pop III and Q3d models are only 12 and 14, respectively. Obviously, the Q3nod case has the most powerful constraint capability among the three MBHB models. The Q3nod model leads to the parameter constraint results $\varepsilon(\Omega_{\rm m})=6.58\%$ and $\varepsilon(h)=1.13\%$ for the $\Lambda$CDM model, and $\varepsilon(w)=20.64\%$ for the $w$CDM model. As a contrast, taking the Q3d model for example, the constraint results, $\varepsilon(\Omega_{\rm m})=9.63\%$ and $\varepsilon(h)=1.79\%$ for the $\Lambda$CDM, and $\varepsilon(w)=33.45\%$ for the $w$CDM, are significantly worse than the Q3nod case.

In Fig.~\ref{TQCMB}, we constrain the cosmological models with the CMB (red), TianQin (grey), and CMB+TianQin (blue) data combination to explore the effect on breaking the degeneracies inherent in CMB with the TianQin (Q3nod) mock data. We show the two-dimensional posterior contours in the $\Omega_{\rm m}$--$H_{0}$ plane for the $\Lambda$CDM model (a), in the $\Omega_{\rm m}$--$H_{0}$ plane for the $w$CDM model (b), and in the $\Omega_{\rm m}$--$w$ plane for the $w$CDM model (c).
We find that, despite the fact that the parameter constraints from TianQin are rather looser than those from CMB, TianQin could break the degeneracies inherent in CMB to some extent because of the apparent
discrepancies between the parameter degeneracy orientations, leading to a great improvement on the constraint accuracies. For the $\Lambda$CDM model, the constraints on $\Omega_{\rm m}$ and $h$ are improved by about 33\% and 32\%, respectively. For the $w$CDM model,
the constraints on $\Omega_{\rm m}$, $h$, and $w$ are improved by about 90\%, 90\%, and 85\%, respectively.

Furthermore, we wish to investigate whether the TianQin's standard siren observation can provide useful help in improving the cosmological estimation under the current mainstream electromagnetic observations. We thus use the TianQin, CMB+BAO+SN, and CMB+BAO+SN +TianQin data combinations to constrain the cosmological models in order. The constraint results are displayed in Fig.~\ref{TQCBS}. In this figure, we exhibit the posterior contours in the $\Omega_{\rm m}$--$H_{0}$ plane for the $\Lambda$CDM (a) and in the $w$--$H_{0}$ plane for the $w$CDM (b). Notably, for the TianQin mock data, we only choose the results simulated in light of the Q3nod model.

We clearly see that, the orientations of the parameter degeneracies formed by CMB+BAO+SN and TianQin are rather different, which implies that the addition of TianQin can effectively break the parameter degeneracies. Indeed, when including the TianQin(Q3nod) mock data into the CBS data sets, the data combination of CMB+BAO+SN +TianQin(Q3nod) gives the results $\varepsilon(\Omega_{\rm m})=1.55\%$, $\varepsilon(h)=0.51\%$ for $\Lambda$CDM, and $\varepsilon(w)=3.04\%$ for $w$CDM, showing that the constraints on $\Omega_{\rm m}$ and $h$ (in $\Lambda$CDM) and $w$ (in $w$CDM) are improved by about 22\%, 24\%, and 20\%, respectively. Further detailed results can be directly found in Tables~\ref{tab:LCDM} and \ref{tab:wCDM}.

Finally, based on the method presented in this paper, we also estimate the ability of LISA to constrain cosmological parameters for a brief comparison. Taking the pop III model as an example, the constraint results, $\varepsilon(\Omega_{\rm m})=6.92\%$ and $\varepsilon(h)=1.40\%$ (in $\Lambda$CDM), and $\varepsilon(w)=25.21\%$ (in $w$CDM), provided by LISA are better than those of TianQin. In truth, the longer arm length gives LISA the advantage over TianQin in detecting MBHBs. Nevertheless, TianQin may have more potential to detect intermediate-mass black holes due to its higher sensitivity at high frequencies.

For the simulation of SNR, we only consider the contribution from the inspiral phase of MBHB coalescence. In fact, for a MBHB emitting gravitational waves at lower frequency, the merger and ringdown phases may also be within TianQin's detection band. Nevertheless, the inspiral-merger-ringdown (IMR) waveform models for efficient parameter estimation are still under development \cite{Bini:2015bfb,Hannam:2013oca,Ossokine:2015vda}. The impact of including the merger and ringdown phases on parameter estimation is discussed in several papers \cite{Babak:2008bu,McWilliams:2011zs,Littenberg:2012uj}. Klein \emph{et al.} \cite{Klein:2015hvg} use the combination of spin-aligned PhenomC IMR waveforms \cite{Santamaria:2010yb} and a restricted set of dedicated precessing IMR hybrid waveforms to estimate the ability of LISA in constraining parameters. They find that $\Delta d_{L}/d_{L}$ in the IMR waveform model would be up to $\sim$ 20 times smaller than that in the inspiral-only model, depending on the mass of MBHB. Tamanini \emph{et al.} \cite{Tamanini:2016zlh} show how the IMR models improve the estimation accuracy of cosmological parameters for the LISA project, and thus we also expect a similar improvement for TianQin. However, for the results of CMB+TQ and CBS+TQ, the accuracy improvement is mainly due to the breaking of degeneracies between cosmological parameters, so the waveform models have a relatively small effect on these results.

It should be added that, in principle, the predicted number of GW events with accompanying EM counterparts originated from the MBHB coalescences should be precisely determined according to the detection capability of future radio/optical projects such as SKA, LSST, and ELT, etc.
However, the certain formation mechanism of MBHB is still unclear up to now, thus for the purpose of merely a preliminary cosmological parameter estimation, we refer to the results provided in Ref.~\cite{Tamanini:2016zlh}, as mentioned in the section~\ref{subsubsection:A2}.
It should be emphasized that, in this study we consider the twin-constellation scenario for TianQin's data simulation, which is actually an optimistic scenario. The capability of TianQin with a single-constellation in constraining cosmological parameters will become relatively weak.

Recently, it has been shown that the future GW standard siren observation from the third-generation ground-based GW detector, the Einstein Telescope (ET), can significantly improve the constraints on numerous cosmological parameters by effectively breaking the parameter degeneracies \cite{Zhang:2019ylr,Zhang:2018byx,Wang:2018lun,Zhang:2019ple}. Of course, for the case of ground-based GW detector, almost all the standard sirens are provided by the BNS merger events at relatively low redshifts, and the number of events is expected to be very large, e.g., about 1000 data are usually produced and used in the simulation of GW standard sirens from the ET. For the case of space-based GW observatory, we see that the event numbers in several various MBHB models are all rather small. Notwithstanding, it is found that, even for the standard siren observation from TianQin, it can also provide important improvements in the future cosmological parameter estimation.

Looking forward, space-based GW detectors will accomplish much in other aspects of cosmology. For example, LISA's observational data can provide help in constraining the early and interacting dark energy models \cite{Caprini:2016qxs}, and can also be used to reconstruct the dark sector interaction \cite{Cai:2017yww}. By using LISA to detect the GW source's luminosity distance, signals that indicate the presence of modified gravity may be detected \cite{Belgacem:2019pkk}. We thus also expect TianQin to have the similar potential in these aspects. In addition, LISA, Taiji, and TianQin may form a detection network in the future \cite{Ruan:2020smc,Wang:2020vkg}, which could further improve the detection rate, sky localization, and parameter estimation accuracy.

\section{Conclusion}\label{sec:con}

TianQin is a space-based GW observatory scheduled to be launched in the 2030s. In this work, we make a preliminary forecast for the cosmological parameter estimation with the GW standard siren observation from TianQin. We simulate the GW standard siren observational data of TianQin based on its 5-year operation after the completion of construction. We consider three models for the population of MBHB, i.e., the pop III model, the Q3nod model, and the Q3d model, to predict the event numbers of MBHB mergers in the simulation. From this investigation, we wish to know: (i) To what extent can the TianQin-only data constrain the cosmological parameters; (ii) To what extent can the TianQin mock data break the parameter degeneracies originated from the CMB observation; (iii) What role would the TianQin's standard siren observation play in the future cosmological parameter estimation.

We consider two simple cosmological models, i.e., the $\Lambda$CDM model and the $w$CDM model, in this work. To investigate the capability of standard siren observation from TianQin in breaking the parameter degeneracies inherent in the CMB observation, we separately use the TianQin's simulated data (based on the MBHB models of pop III, Q3nod, and Q3d), the CMB data, and the CMB+TianQin data to constrain the cosmological models. In addition, in order to explore the TianQin's capability in improving the parameter constraint accuracies under current mainstream observations, we further place the constraints on the cosmological models under the CMB+BAO+SN, and the CMB+BAO+SN+TianQin data combinations.

For the TianQin-only case, we find that the simulated data based on the Q3nod model can provide the tightest constraints on the cosmological parameters among all the three MBHB models, with the constraint results being $\varepsilon(\Omega_{\rm m})=6.58\%$ and $\varepsilon(h)=1.13\%$ for $\Lambda$CDM, and $\varepsilon(w)=20.64\%$ for $w$CDM.

For the CMB and TianQin data, their constraints on the cosmological parameters are quite different in the orientations of parameter degeneracies. Especially for the $w$CDM model, the parameter degeneracies inherent in CMB are completely broken, and the precision of constraints is improved greatly. Concretely, when adding the TianQin(Q3nod) mock data into the CMB observation, the constraint results for the $w$CDM become $\varepsilon(\Omega_{\rm m})=2.02\%$, $\varepsilon(h)=1.05\%$, and $\varepsilon(w)=3.64\%$, and the constraint precision on the parameters $\Omega_{\rm m}$, $h$, and $w$ would be improved by about 90\%, 90\%, and 85\%, respectively.


Finally, when adding the TianQin(Q3nod) mock data into the current optical data sets (i.e., the CMB+BAO+SN data combination), we have the results $\varepsilon(\Omega_{\rm m})=1.74\%$ and $\varepsilon(h)=0.57\%$ for the $\Lambda$CDM, and $\varepsilon(w)=3.73\%$ for the $w$CDM, which implies that with the addition of TianQin mock data, the constraints on $\Omega_{\rm m}$ and $h$ (in $\Lambda$CDM) and $w$ (in $w$CDM) could be improved by about 26\%, 29\%, and 13\%, respectively. Therefore, we can conclude that the standard siren observation from TianQin would provide important improvements in the cosmological parameter estimation in the future.

\begin{acknowledgments}
We are very grateful to Enrico Barausse, Nicola Tamanini, Dong-Ze He, Yi-Ming Hu, Jian-Wei Mei, Shao-Jiang Wang, and Tao Yang for fruitful discussions. This work was supported by the National Natural Science Foundation of China (Grants Nos. 11975072, 11835009, 11875102, 11690021, and 11522540), the Liaoning Revitalization Talents Program (Grant No. XLYC1905011), the Fundamental Research Funds for the Central Universities (Grant No. N2005030), and the National Program for Support of Top-Notch Young Professionals (Grant No. W02070050).
\end{acknowledgments}

\end{document}